\documentclass[aps,prd,a4paper,superscriptaddress,nofootinbib,showpacs,showkeys,amsfonts,amssymb,amsmath]{revtex4}
\usepackage{amssymb,latexsym}
\usepackage{amsmath, amsthm}

\usepackage{amscd}
\usepackage{times}
\usepackage{epsfig}
\usepackage{psfrag}
\usepackage{graphicx}

\begin{document}
\title[]{All black holes in Lema\^{i}tre-Tolman-Bondi inhomogeneous dust collapse}

\author{Pankaj S. Joshi}
\email{psj@tifr.res.in}
\affiliation{Tata Institute of Fundamental Research, Homi Bhabha Road, Colaba, Mumbai 400005, India}

\author{Daniele Malafarina}
\email{daniele.malafarina@nu.edu.kz}
\affiliation{Center for Field Theory and Particle Physics \& Department of Physics,
Fudan University, 220 Handan Road, 200433 Shanghai, China}
\affiliation{​Physics Department, SST, Nazarbayev University, 53 Kabanbay Batyr avenue,
010000, Astana, Kazakhstan}

\swapnumbers
\begin{abstract}
Within the Lema\^{i}tre-Tolman-Bondi formalism for gravitational collapse of inhomogeneous
dust we analyze the parameter space that leads to the formation of a globally covered singularity (i.e. a black hole) when some physically reasonable requirements are imposed (namely positive radially decreasing and quadratic profile for the energy density and avoidance of shell crossing singularities).
It turns out that a black hole can occur as the endstate of collapse only if the singularity is simultaneous as in the standard Oppenheimer-Snyder scenario.
Given a fixed density profile then there is one velocity profile for the infalling particles that will produce a black hole. All other allowed velocity profiles will terminate the collapse in a locally naked singularity.
\end{abstract}


\maketitle

\section{Introduction}

The first gravitational collapse model to be studied thoroughly within the general theory of relativity was the very well known Oppenheimer-Snyder (OS) homogeneous dust collapse
\cite{OS}.
As it is known the OS model terminates in a simultaneous singularity which is covered by an event horizon. Therefore the main elements characterizing the OS model from the metric side are two, namely the occurrence of a simultaneous singularity and the appearance of trapped surfaces  before the singularity and they are closely linked to the assumption of homogeneity for the density profile.

As more general matter profiles started to be investigated it became clear that the OS model was not the only possible final fate for the complete collapse of a spherical matter cloud and that a simultaneous singularity is a fine tuned feature appearing only in certain collapse scenarios.
The simplest generalization of the OS model is the well known Lema\^{i}tre-Tolman-Bondi (LTB) inhomogeneous dust collapse
\cite{LTB}.
As inhomogeneities are introduced in the pressureless matter profile the simultaneous singularity structure is lost. Different shells become singular at different times. Not only the simultaneity of the singularity changes but also, and more importantly, the behaviour of the horizon is affected leaving open the possibility for the singularity developing at the central shell to be locally or globally naked.
In fact as it turns out some matter profiles still present the horizon forming before the formation of the singularity while others develop trapped surfaces at the time of formation of the singularity therefore leaving the possibility for geodesics to come out of the ultra high density region that develops at the center of the cloud. As it has been shown by many authors over the past decades the singularity that develops in these collapse models is naked, at least locally
\cite{dust}.
Mathematically these models serve as counterexamples to some formulation of the Cosmic Censorship Conjecture (CCC)
\cite{Penrose},
which states that every physically reasonable collapse process must generically lead to the formation of singularities that are covered by a horizon at all times.
On the other hand the issue is important also from an astrophysical point of view because the possible existence of naked singularities means that the regions of extremely high densities where classical general relativity breaks down can be casually connected to the outside universe and therefore bear an observational signature. Of course the LTB models (and the OS model which is a subcase of LTB) are idealized mathematical models that do not describe a realistic star. Still, the issue of visibility or otherwise of naked singularities is physically very important as the OS model and CCC are at the foundation of all of black hole physics which is used in astrophysical applications today. These simple models then provide great insights in the important elements that determine the final fate of collapse.

 In more recent times several classical gravitational collapse scenarios
 have been studied, with many different matter models.
 The picture that emerged is that under general conditions to ensure the
physical validity of the matter models both black holes and naked singularities
can arise as the endstate of collapse (see \cite{review} and references therein for a recent review).
Many examples have been found that lead collapse to the formation
 of a naked singularities even when pressures are allowed
\cite{ns}.
Furthermore these scenarios seem to be sufficiently generic
(once a suitable definition of `genericity' is given in this context)
and stable with respect to small perturbations in the initial data
\cite{genericity}.
Therefore it has become essential to isolate the conditions
under which a physically realistic collapse will go to a black
hole, developing from regular initial data.

This is also necessary in view of the increasing amount of astrophysical
applications of black holes and in view of the absence of a proof for the
CCC, which is fundamental to black hole physics.
If we assume that singularities must be resolved within a theory of quantum gravity, then classical solutions with naked singularities might be considered as a theoretical window open on new physics in astrophysical phenomena
(see for example \cite{QG} for approaches based on Loop Quantum Gravity).
Therefore, given the lack of a viable general proof of the CCC and the increasing amount of theoretical evidence in favour of the naked singularities in recent times researchers have begun to consider the observational features that these solutions might bring
(see for example \cite{obs}).


In the present article we look for a characterization of the black hole formation process in a well know model, namely the LTB collapse scenario, when a series of basic physical requirements, such as the positivity and decreasing radial behaviour of the density, quadratic behaviour near the origin of the density and the absence of shell crossing singularities, are imposed.
We show that once we impose the above conditions the only models developing a black hole as the final fate of collapse are those for which the singularity is simultaneous. All other allowed scenarios having a non-constant singularity curve develop a locally naked singularity at the center of the cloud.

In section \ref{LTB} we review the Lema\^{i}tre-Tolman-Bondi scenario and outline its main features while in section \ref{allBH} we describe the conditions for no shell crossing and see how from these we can characterize entirely the possible outcomes of collapse. In section \ref{simul} we derive explicitly the parameter space that leads to the formation of black holes with a simultaneous singularity.
Finally in section \ref{conc} we discuss the results and their possible implications for astrophysics.
In the following we use units in which $G=c=1$ and for simplicity we absorb the factor $k=8\pi G$ that appears in Einstein's equations in the energy momentum tensor.

\section{Lema\^{i}tre-Tolman-Bondi models}\label{LTB}

The Lema\^{i}tre-Tolman-Bondi metric describing inhomogeneous dust in comoving coordinates is given by
\begin{equation}
    ds^2=-dt^2+\frac{R'^2}{1+f}dr^2+R^2d\Omega^2 \; ,
\end{equation}
where $R=R(r,t)$ and $f=f(r)$. The energy momentum tensor takes diagonal form and is given by $T^0_0=\rho$, $T^i_i=p=0$ (with $i=1, 2, 3$).
Then Einstein's equations simply reduce to
\begin{equation}
  \label{rho}
  \rho=\frac{F'}{R^2R'} \; , \; \;
  p =-\frac{\dot{F}}{R^2\dot{R}}=0 \; ,
\end{equation}
where $(')$ denotes derivatives with respect to $r$ and $(\dot{})$ denotes derivatives with respect to $t$.
Requiring the metric to be lorentzian imposes a condition on the energy function $f(r)$, namely $f\geq -1$, while
the function $F$, called the Misner-Sharp mass, describing the amount of matter enclosed by the shell
labeled by $r$, is required to be non negative and radially increasing and it is given by $1-F/R=g_{\mu\nu}\nabla^\mu R\nabla^\nu R$.
From the second of Eqs.~\eqref{rho} we see immediately that we must have $F=F(r)$,
which means that the amount of matter enclosed in any shell labeled by $r$ is conserved throughout collapse.
The Misner-Sharp mass equation
then can be rewritten in the form of an equation of motion as
\begin{equation}\label{motion}
    \dot{R}=\pm\sqrt{\frac{F}{R}+f} \; ,
\end{equation}
with the plus sign to describe expansion and the minus sign to describe collapse.
In the following we will consider the case of collapse. In general $F$ and $f$ are free parameters of the system and they must be chosen in order to satisfy the physical validity of the model. The solution obtained from the integration of the above equation can always be matched with a Schwarzschild exterior at a boundary surface $R_b(t)=R(r_b,t)$
\cite{matching}.

The general solution of Eq.~\eqref{motion} has been studied thoroughly for both collapse and cosmological models
(see for example \cite{cosmology}).
It is easily shown that there are three different cases depending on the sign of $f$. The hyperbolic case, given by $f>0$, corresponds to unbound collapse. The particles in the cloud have positive initial velocity in the limit as $R$ goes to infinity. The flat case, given by $f=0$, corresponds to marginally bound collapse. The particles in the cloud have zero initial velocity in the limit as $R$ goes to infinity. The elliptic case, given by $f<0$, corresponds to bound collapse. The argument under the square root in Eq. \eqref{motion} becomes zero at a finite $R$ and the particles in the cloud have negative initial velocity in the limit as $R$ goes to infinity.

In general given a curve $R_{\gamma}(r)$ we will have $t_{\gamma}(r)=t(r,R_{\gamma}(r))$, from which we get the general expression
$dt_{\gamma}/dr=\partial t/\partial r+(\partial t/\partial R)(d R_{\gamma}/dr)$.
The curves that are most relevant for the study of the solutions of Eq. \eqref{motion} in gravitational collapse are:
(1) The singularity curve, given by $R_s(r)=0$. Then $t_s(r)=t(r,0)$ and
        $t_s'=\left(\partial t/\partial r\right)_{R=0}$. It describes the time at which the shell labelled by $r$ becomes singular.
(2) The apparent horizon, given by $R_{ah}(r)=F(r)$. Then $t_{ah}(r)=t(r,F(r))$ and
        $t_{ah}'=\left[\partial t/\partial r+(\partial t/\partial R)F'\right]_{R=F}$. It describes the time at which the shell labelled by $r$ becomes trapped.
(3) The shell crossing curve, given by $R_{sc}'(r)=0$. Then $t_{sc}(r)$ is given by $R'(r,t_{sc}(r))=0$ and it describes the time at which the shell labelled by $r$ intersects another shell signaling the breakdown of the coordinate system.

The singularity curve does not strictly belong to the manifold and can be considered as the `boundary' of the spacetime as no other curve can be prolonged past it. In dust collapse it indicates the presence of a strong curvature singularity and physical quantities such as the energy density $\rho$ diverge along the curve.
The apparent horizon curve is the boundary of the region of formation of trapped surfaces. As in the Schwarzschild case it is given by the condition that the surface $R(r,t)={\rm const.}$ becomes null, which translates to $g^{\mu\nu}\partial R_\mu\partial R_\nu=0$.
The shell crossing curve also indicates the presence of a singularity, as can be seen by the first of Eq.~\eqref{rho}, but in this case it is a weak curvature singularity and the spacetime can be extended through it.
Finally another crucial element for the global features of the spacetime is the boundary curve, given by $r=r_b$ that corresponds to a shrinking area-radius $R_b(t)=R(r_b,t)$. Given the absence of pressures, in the dust models it is always possible to choose the boundary at will.
In the case of dust collapse, and in cases with pressure where the final central singularity is massive, the singularity curve is spacelike and the only portion that can be visible to far away observers is the center,
namely $t_s(0)$. Nevertheless it is possible to construct models of collapse of perfect fluids where the singularity curve becomes timelike and is uncovered for longer times
\cite{JDM}.


We shall take the initial time $t_i=0$ such that $R(r,t_i)=r$. This is always possible due to the scaling
degree of freedom left for $R$. Then we can introduce a scaling function $v(r,t)$ defined by
 $R(r,t)=rv(r,t)$,
with the initial condition $v(r,0)=1$.
In order to be physically reasonable the matter cloud must satisfy certain requirements, such as regularity of the density at the center during collapse before the formation of the singularity and weak energy condition. Regularity of $\rho$ at $r=0$ at the initial time imposes that
\begin{equation}\label{M}
    F(r)=r^3M(r) \; , \; \; \text{and} \; \; f = r^2b(r) \; ,
\end{equation}
with $M$ a positive function. The free functions $M$ and $b$ can be thought of as describing the denisty inhomogeneities and the velocity profile of the particles in the cloud. Their physical interpretation in connection with inhomogeneous cosomological models and collapse models have been thoroghly studied (see for example
\cite{Sussman} for a characterization in terms of entropic favorable models).
With the above substitution the equation of motion becomes
\begin{equation}\label{motion2}
    \dot{v}=-\sqrt{\frac{M}{v}+b} \; .
\end{equation}
A further initial condition for collapse to occur is then given by $b+M\geq 0$, which must be added to the condition for the metric to be Lorentzian given by $b\geq -1/r^2$ and constraints the allowed functions $b$ in the elliptic case.

For the model to be physically reasonable we require $\rho$ to be positive, and therefore satisfying the weak energy condition, and radially non increasing outwards.
The condition that $\rho$ be positive is achieved when $F'>0$ and $R'>0$. Since we require $M(0)>0$ it is easy to check that the case $F'<0$ and $R'<0$, that would also give a positive density, is not allowed because it would imply $M<0$ near the center.
Therefore to have $\rho> 0$ and finite we must require the two conditions
\begin{eqnarray}
  3M &>& -rM'   \; , \\
  R' &>& 0 \; .
\label{condR'}
\end{eqnarray}
From the first one we see that we must have $M(0)=M_0>0$, while the second condition implies the avoidance of shell crossing singularities.
The second physical requirement on $\rho$ is that the energy density be a non increasing function of $r$. This is achieved if $\rho'\leq 0$, which gives the further condition
\begin{equation}\label{rho'}
    F''\leq F'\left(\frac{2R'}{R}+\frac{R''}{R'}\right) \; .
\end{equation}
Typically the energy density is chosen in such a way that it can be written as a power series close to $r=0$ as
\begin{equation}
    \rho=\rho_0(t)+\rho_1(t)r+o(r^2) \; ,
\end{equation}
where we have $\rho_0(t)=3M_0/v(0,t)^3$ and $\rho_1(t)=4M'(0)/v(0,t)^3-12M_0v'(0,t)/v(0,t)^4$. At the initial time, for which $v=1$ and $v'=0$, these become $\rho_0(0)=3M_0$ and $\rho_1(0)=4M'(0)$. From this we see that having $\rho$ non increasing radially implies that $M'(0)\leq 0$.
If we add the further requirement that only quadratic terms in $r$ appear in the expansion, as it is done in most models of astrophysical interest, we obtain that $M'(0)=0$, in agreement with the usual requirement that $\rho$ have no cusps at the origin, and conclusions similar to the ones above must be drawn for $M''(0)$.
In the following we shall consider mass profiles $M$ and velocity profiles $b$ that can be expressed as a polynomial near $r=0$. Imposing that the energy density has only quadratic terms in $r$ implies that we must choose
\begin{eqnarray}\label{M}
M(r)&=&M_0+M_2r^2+o(r^3) \; , \\ \label{b}
b(r)&=&b_0+b_2r^2+o(r^3) \; .
\end{eqnarray}
Integrating fully Eq.~\eqref{motion2} one must consider the three cases separately. In the flat region given by $b=0$ we get
        \begin{equation}
           t(r,v)= -\frac{2v^{\frac{3}{2}}}{3\sqrt{M}}+t_s(r) \; ,
        \end{equation}
and the singularity curve $t_s(r)=t(r,0)$ is given by imposing the initial condition $v(r,0)=1$. The corresponding solutions in the hyperbolic and elliptic region are similarly obtained (see the Appendix for details).
Note that the functional dependence of $t$ does not change wether we consider the coordinates $(r,v)$ or $(r,R)$ (area-radius coordinates), provided that we change $b$ with $f$ and $M$ with $F$. The same holds true for $R'$, given below, and $\dot{R}$, namely they both show the same functional dependence in terms of $f$ and $F$ as $v'$ and $\dot{v}$ do in terms of $b$ and $M$.

\section{All black holes} \label{allBH}

By black hole we mean a solution of Eq. \eqref{motion2} for which the shell focusing singularity is hidden behind the apparent horizon at all times.
The Kretschmann scalar for the LTB metric is
\begin{equation}
    \mathrm{K}=\frac{12F^2}{R^6}+\frac{8FF'}{R^5R'}+\frac{3F'^2}{R^4R'^2} \; ,
\end{equation}
from which we see that the metric becomes singular at $R=0$ and also at $R'=0$. As said before the condition $R'=0$ denotes the presence of a shell crossing
singularity. These were the first `naked singularities' to be studied in collapse models
\cite{cross}.
Typically shell crossing singularities in LTB are `weak', in the sense that they are due to
caustics arising when different shells overlap and the spacetime be extended through them
\cite{cross2}.

Note that not always $R'=0$ implies a shell crossing singularity. In fact if we have $F'=0$ in such a way that $F'/R'$ is finite as $R'$ goes to zero
the Kretschmann scalar remains finite as well and we thus have a so-called `neck'.
Nevertheless in collapse models we typically deal with functions $F$ that are monotonic, therefore ruling out this case which can be relevant in cosmological models
\cite{cosmology}.
The condition for avoidance of shell crossing is then given by Eq.~\eqref{condR'}.
Once we solve the equation of motion to obtain $t(r,R)$ we can evaluate $R'=-(\partial t/\partial r)\dot{R}$,
from which we can already see that for collapse, if we require no shell crossing, we must have $\partial t/\partial r>0$.
After some calculations we get
\begin{equation}\label{R'}
    R'=\left(\frac{F'}{F}-\frac{f'}{f}\right)R-\left[t_s'+\left(\frac{F'}{F}-\frac{3f'}{2f}\right)(t-t_s)\right]\sqrt{\frac{F}{R}+f} \; ,
\end{equation}
and shell crossing singularities can be avoided provided that
$R'>0$.
Let us now focus on the marginally bound case for the sake of clarity. Eq.~\eqref{R'} becomes
\begin{equation}
    R'=\frac{1}{3}\frac{F'}{F}\frac{R^{\frac{3}{2}}-r^{\frac{3}{2}}}{\sqrt{R}} \
    +\frac{\sqrt{r}}{\sqrt{R}} \; ,
\end{equation}
from which we see that in this case, imposing the condition for no shell crossing implies
\begin{equation}
    3F>F'\left(r-\frac{R^{\frac{3}{2}}}{\sqrt{r}}\right) \; \Leftrightarrow \; M'(1-v^{\frac{3}{2}})<0 \; ,
\end{equation}
which, since $v\in[0,1]$, in turn implies $M'<0$.
We can write the shell crossing curve as
\begin{equation}
    t_{sc}(r)=\frac{2\sqrt{M}}{3M+rM'} \; ,
\end{equation}
and it is easy to see that if $M={\rm const}.$ then $t_{sc}=t_s$ while if $M'<0$ then $t_{sc}\geq t_s$, with the equal sign holding only at $r=0$, and no shell crossing occur in the spacetime.
On the other hand the singularity curve is given by $t(r,0)$ and the condition that the singularity curve is non increasing is given by
\begin{equation}
    t_s'= \frac{\sqrt{r}}{\sqrt{F}}\left(1-\frac{1}{3}\frac{F'r}{F}\right)\leq 0 \; ,
\end{equation}
which corresponds to
\begin{equation}
    3F\leq F'r \; \Leftrightarrow \; M'\geq0 \; .
\end{equation}

Now for the matter profile given by Eq.~\eqref{M} with $M_2\neq 0$ the condition for avoidance of shell crossing singularities translates to $M_2<0$. On the other hand it has been shown
(see \cite{dust})
that in this case collapse leads to a locally naked singularity\footnote{In the case where $M_2=0$ one has to consider the next order. Then it can be shown that for $M_3\neq 0$ there exist a limiting value for $M_3$ above which collapse leads to a locally naked singularity, while when the first non zero term is $M_n$ with $n>3$ collapse leads to the formation of a black hole. Nevertheless for `realistic' models is always reasonable to consider $M_2\neq 0$.}.
We therefore see that, with the only exception of simultaneous collapse for which $t_s(r)=t_0$, globally covered singularity and no shell crossing are incompatible conditions in the case of the marginally bound LTB collapse when the energy density has quadratic terms in $r$. On the other hand having the energy density positive and non increasing is compatible with the condition for avoidance of shell crossing singularities.

Going back to the general case, the conditions for avoidance of shell crossing were given by Hellaby and Lake
\cite{Hellaby}.
Assuming that $F$ is a positive increasing function, $F' > 0$, they can be written as follows
\begin{enumerate}
  \item Flat region ($f=0$):
  \begin{eqnarray}
    t_s' &\geq& 0 \; .
  \end{eqnarray}
  \item Hyperbolic region ($f>0$):
  \begin{eqnarray}
    t_s' &\geq& 0  \; , \\
    f'&\geq&0 \; .
  \end{eqnarray}
  \item Elliptic region ($f<0$):
    \begin{eqnarray}
    t_s' &\geq& 0  \; , \\
    \frac{F'}{F}- \frac{3f'}{2f} &\geq&\frac{2}{3}\frac{t_s'}{F}(-f)^{\frac{3}{2}} \; .
  \end{eqnarray}
\end{enumerate}
We see that in all three cases the singularity curve $t_s(r)$ must be either constant or increasing.
It is easy to see that close to the center the apparent horizon behaves like the singularity curve.
In fact from $t(r,R)$ the apparent horizon curve in the flat region is given by $t_{ah}(r)=t(r,F(r))$ which corresponds to
  \begin{equation} \label{ah-flat}
  t_{ah}(r)= t_s(r)-\frac{2}{3}F(r) \; .
  \end{equation}
 Therefore, for the matter profile under consideration here, where $F(r)=r^3M(r)$ with $M$ given by Eq. \eqref{M} with $M_2\neq 0$,  we see that $F'\simeq r^2$ while $t_s' \simeq r$. Then in this case from the requirement that $t_s'>0$ there will always be a finite neighborhood of $r=0$ in which $t_s'>2F'/3$, which implies that $t_{ah}$ is increasing near $r=0$.
The apparent horizon in the other cases is easily calculated in the same way (see Appendix for details).
In all three cases we have $t_{ah}(r)= t_s(r)+o(r^2)$ for $r$ close to zero. It can be shown that under sufficiently general circumstances, like those obtained by matter and velocity profiles as in Eqs. \eqref{M} and \eqref{b}, an increasing apparent horizon is a sufficient condition for the local visibility of the central singularity
(see for example \cite{dust} or \cite{Giambo} for a strictly mathematical treatment).
In fact the equation for outgoing radial null geodesics $t_{\gamma}(r)$ can be written as $dt_{\gamma}/dr=R'/\sqrt{1+f}$, and the analisys close to the center can be done in the elliptic and hyperbolic cases following the same strategy as in the flat case.
This shows that the singularity must be locally naked in the allowed cases when $M_2<0$.

 From the above we understand that imposing avoidance of shell crossing singularities implies that the only case in which the singularity curve can be trapped at all times is given by $t_s(r)=t_0$.
Therefore we conclude that if we require physical reasonable profiles for density and velocity profiles, like those given by Eqs.~\eqref{M} and \eqref{b},
the only case where a black hole (where we adopt the strict definition of black hole meaning a solution where the singularity is neither globally nor locally naked) can form from the complete collapse of inhomogeneous dust is that of simultaneous collapse. Other physically valid configurations with $M_2\neq 0$ will lead the central singularity forming as the endstate of collapse to be, at least locally, visible.


From the arguments above we see that for a matter profile as in Eq.~\eqref{M} a locally
naked singularity will form `generically' at the end of collapse when the singularity
curve is not simultaneous. On the other hand global visibility is an entirely different matter and some considerations on global visibility are in order here.
Since in principle there could be matter profiles for which the apparent horizon increases until a certain radius and then decreases thus hiding the singularity to observers at spatial infinity
one cannot say anything about global visibility unless a proper definition of the boundary is provided. This is due to the fact that the higher order terms in the expansions of $M$ and $b$ become increasingly more important as we move away from the center.
Some matter profiles will cause the apparent horizon curve to be increasing close
to the center and decreasing away from the center and this can cause the singularity to
be globally covered, although locally naked. Some other matter profiles,
on the other hand, will have $t_{ah}(r)$ increasing from the center until the
boundary thus leaving a globally naked singularity
\cite{jhingan}.
Nevertheless, as far as we consider here only the mathematical aspects of dust models, it is in principle always possible to choose the boundary of the cloud suitably so that the apparent horizon is strictly increasing and the singularity globally naked.
Conversely if we consider a fixed boundary then the velocity profile
and the matter profile can always be chosen is such a way that
the singularity be globally visible.
A sufficient condition for global visibility is given by
\begin{equation}
    t_{ah}'>0 \; ,
\end{equation}
for $r\leq r_b$. This condition implies a condition on $F$ and $f$. For example in the flat case from Eq.~\eqref{ah-flat} it is easy to see that the sufficient condition for global visibility is
\begin{equation}
    t_s'>\frac{2}{3}F' \; ,
\end{equation}
for $r\in[0,r_b]$. In general the sufficient condition becomes
\begin{equation}
    t_s'>\pm\frac{F}{\sqrt{f+1}}\frac{f'}{f}-\left(\frac{F'}{F}-\frac{3f'}{2f}\right)t_{ah} \; ,
\end{equation}
with the plus sign in the hyperbolic case and the minus sign in the elliptic case.

\section{Simultaneous collapse}\label{simul}

We investigate now the structure of all possible collapse models with quadratic matter profiles that lead to the formation of a globally covered singularity.
We have seen that these correspond only to the case of a simultaneous singularity.
Collapse is simultaneous if all the shells fall into the central shell focusing singularity
at the same comoving time $t_0$. Therefore the necessary and sufficient condition for
simultaneous collapse is $t_s(r)=t_0$.
In the case of marginally bound collapse (corresponding to the flat region) we have
\begin{equation}
    t_s(r)=\frac{2}{3\sqrt{M}} \; ,
\end{equation}
therefore $t_s(r)={\rm const.}$ is satisfied only if $M=M_0$ which corresponds
to the Oppenheimer-Snyder (OS) collapse model. On the other hand from the condition $t_s'(r)>0$
to avoid shell crossing singularities we get
\begin{equation}
    t_s'(r)=-\frac{M'}{3M^{\frac{3}{2}}}\geq 0 \; ,
\end{equation}
which is satisfied for $M'\leq 0$ (with equal sign only at $r=0$). This is in agreement with an energy density profile which is positive
and decreasing and with the formation of a locally naked singularity.

\begin{figure}
 \begin{minipage}[h]{8.5cm}
   \centering
   \includegraphics[width=8cm]{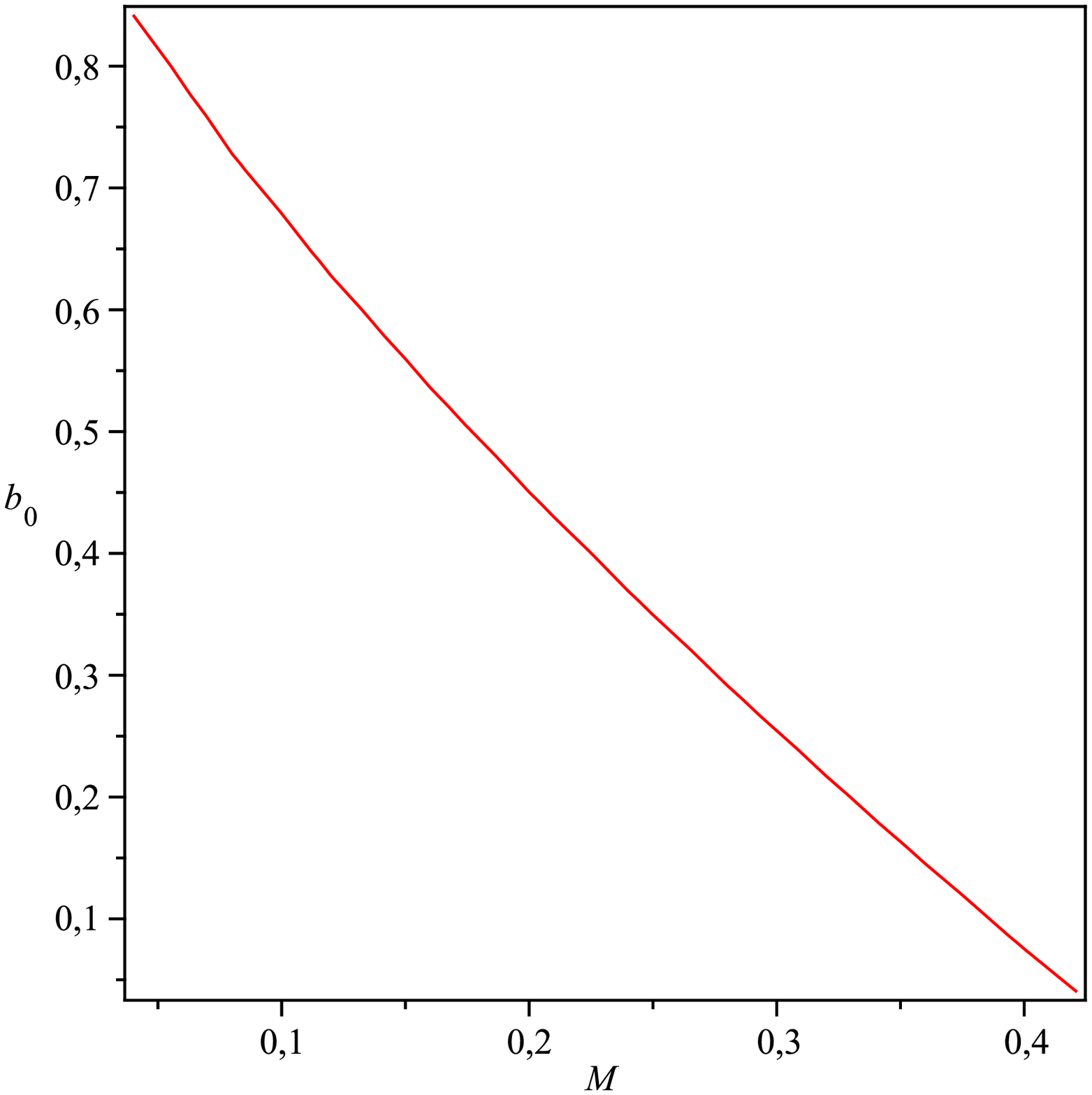}
 \end{minipage}
 \begin{minipage}[h]{8.5cm}
  \centering
   \includegraphics[width=8cm]{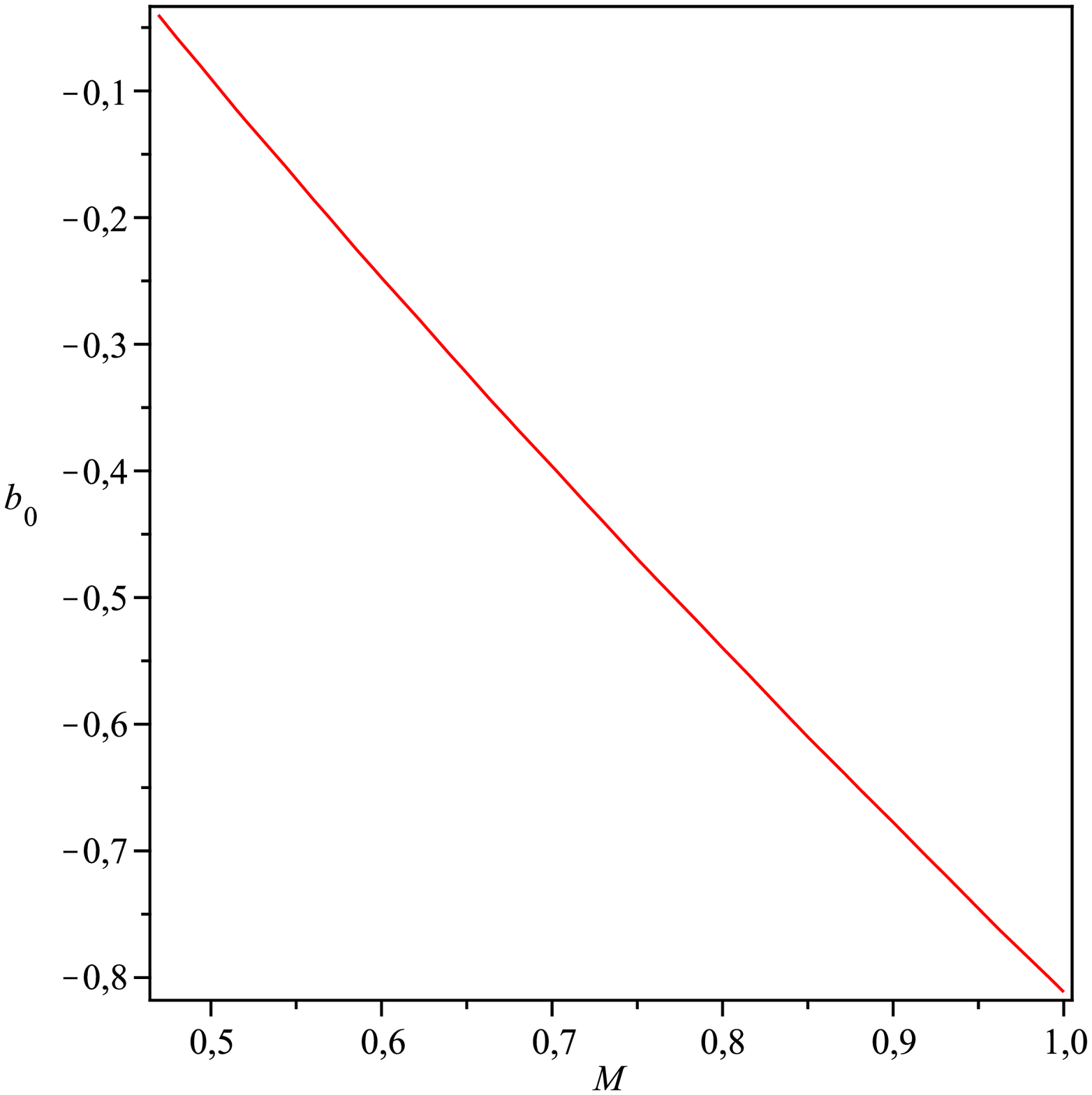}
 \end{minipage}
   \caption{The implicit plot of $b(M)$ from $T(b,M)$ with $t_0= 1$ in the hyperbolic case
   (on the left) and in the elliptic case (on the right).}
\label{fig}
\end{figure}

In the hyperbolic and elliptic cases the singularity curve can be easily obtained (see Appendix).
If we assume $b={\rm const}.$ we can see again that the condition for simultaneous collapse
imposes $M'=0$ and once again we retrieve the OS collapse scenario.
Nevertheless the OS homogeneous dust collapse scenario is not the only case where a simultaneous singularity can be present.
In fact if we write the singularity curve as $t_s(r)=t_s(b(r),M(r))=t_s(b,M)$ then the condition of simultaneous singularity $t_s=t_0$ is satisfied on the zero surfaces of the function $T(b, M)=t_s(b,M)-t_0$. This means that in order to have a black hole we must take $b$ as a function of $M$, given by $b(r)=b(M(r))$, which is implicitly defined by $T(b,M)=0$. This is in general always possible without any loss of generality due to the monotonic behaviour of $M(r)$ and shows that for any given mass profile $M(r)$, provided that $M_2\neq 0$, there will be one velocity profile $b(r)$ which will terminate the collapse in a black hole, while all other possible choices of $b$ that avoid shell crossing will make the collapse terminate in a locally naked singularity.
To evaluate explicitly the velocity profiles $b(r)$ for a simultaneous singularity
when $b\neq {\rm const}.$ we then impose $T(b,M)=0$ with $T$ being at least a $\mathcal{C}^1$
function. We then obtain $b(M)$ from the implicit function theorem. The solution need
not be easily found analytically but it is always possible to evaluate the solution numerically (see Fig. \ref{fig}).

Now we will concentrate on the behaviour of the mass and velocity profiles near the center of the cloud. These parameters represent the density fluctuations and spatial curvature of the spacetime at a given fixed initial time slicing and can be prescribed arbitrarily (for a thorough study of their interpretation see for example
\cite{Sussman2}).

The general formalism developed in
\cite{JG}
to study spherically symmetric gravitational collapse type I matter clouds
can be easily applied to the LTB scenario described above.
It is easy to show that requiring the energy density to be $\mathcal{C}^2$ at
the center implies that the singularity curve must also be $\mathcal{C}^2$ at
the center. Therefore, if we assume that the behaviour near the center of $M$
and $b$ can be written as $M(r) = M_0+M_1r+M_2r^2+o(r^3)$ and
$b(r) = b_{0}+b_{1}r+b_{2}r^2+o(r^3)$,
where for simplicity we have kept also the terms of order one, it follows that we can also expand the singularity curve near the center as
\begin{equation}
    t_s(r)=t_0+\chi_1r+\chi_2r^2+... \; .
\end{equation}
If we impose the mass and velocity profiles to have the behaviour as in Eqs. \eqref{M} and \eqref{b} the same reasoning applies to the second order terms, while the first order terms vanish.
We can explicitly evaluate the coefficients of the expansion that turn out to be
\begin{eqnarray} \label{chi1}
  \chi_1 &=&-\frac{1}{2}\int^1_0\frac{M_1+b_{1}v}{\left(M_0+b_{0}v\right)^{\frac{3}{2}}}\sqrt{v}dv  \; , \\ \label{chi2}
  \chi_2 &=&\frac{3}{8}\int_0^1\frac{(M_1+b_{1}v)^2}{\left(M_0+b_{0}v\right)^{\frac{5}{2}}}\sqrt{v}dv
  -\frac{1}{2}\int^1_0\frac{M_2+b_{2}v}{\left(M_0+b_{0}v\right)^{\frac{3}{2}}}\sqrt{v}dv \; .
\end{eqnarray}
The condition for simultaneous collapse $t_s(r)=t_0$ translates into
$\chi_n=0$ for every $n\geq 1$ and therefore, given a density profile $M(r)$ as
above we see that we must choose every $b_n$ suitably in order to have
$\chi_n=0$. In fact, once we choose $b_{0}$ and impose $\chi_1=0$ we can
obtain $b_1$ from Eq. \eqref{chi1} as
\begin{equation}
    b_1=-\frac{\alpha_1}{\beta_1} \; ,
\end{equation}
with
\begin{equation}
\alpha_1=\int_0^1\frac{M_1\sqrt{v}}{\left(M_0+b_{0}v\right)^{3/2}}dv \; \; \text{and} \; \; \beta_1=\int_0^1\frac{v^{3/2}}{\left(M_0+b_{0}v\right)^{3/2}}dv \; .
\end{equation}
Similarly
from Eq. \eqref{chi2} we see that once $b_{1}$ is given from the above there will be
one $b_{2}$ for which $\chi_2=0$. The same reasoning applies to every order.
Then the velocity profile $b$ which gives rise to a collapse ending in a
black hole will be given by
\begin{equation}
    b(r)=\sum_{n=0}^{+\infty} \frac{b_n}{n!}r^n \; .
\end{equation}
Even if the matter profile's expansion is truncated at some order $N$
(which means that the density profile will be truncated at the same order)
we see that the velocity profile that gives rise to a simultaneous singularity
can always be written as a series with all terms of order $n>N$ equal to zero.
From the above considerations we see also that the time $t_0$ at which the
singularity occurs is determined by the
zeroth order of the mass profile $M_0$ and the velocity profile $b_{0}$.
Finally we note here that the terms of the expansions are all correlated so that
if we require the density to have only even terms in $r$ (as it is often done in
astrophysical models) by assuming that $\rho_{2n+1}=0$ for all $n$, then it
will follow that also $M$, $b$ and $t_s$ will all have only even terms.

\section{Concluding remarks}\label{conc}

We have considered here the final outcomes for a widely studied
class of models describing collapse of inhomogeneous dust clouds.
It is known that depending on the initial configuration these can be
characterized as having singularities that are either globally covered
(i.e. black holes) or naked (locally or globally).
In the case of a black hole final outcome the trapped surfaces form at a
time anteceding the formation of the singularity, while in the naked singularity
case the central shell becomes trapped at the time of formation of the singularity.
This means that null geodesics can propagate from the central singularity
to reach far away observers
\cite{Geo}.
We investigated the conditions for the formation of black holes once
some crucial physical requirements are imposed.
Namely we required that the energy density be positive,
radially decreasing with a quadratic polynomial expansion near the center, we require the velocity profile to satisfy usual regularity conditions and to be small and finally we require that
no shell crossing singularities occur at any time.
Shell crossing singularities have been widely studied in the context of
inhomogeneous cosmological models (see for example
\cite{cosmology}),
where the density profiles can have many different forms.
In the gravitational collapse models considered here on the other hand one must
simply require that the density be non increasing in the outward radial direction.
We showed that under these
circumstances a black hole can form only when
all shells become singular at the same comoving time.
In the case of marginally bound collapse this corresponds to the
requirement that the energy density be homogeneous.
We have shown also that for any given density profile,
this condition implies a specific choice of the velocity profile
of the particles in the cloud.

This analysis provides some insight on the genericity of black hole and naked singularity formation.
In fact we have shown that, at least in the simple case of dust, it would seem that (at least locally)
naked singularities are a generic outcome once some realistic conditions are required.
Of course if we understand the surroundings of the singularity as a region of very high
density where the classical relativistic model breaks down there is no reason
to exclude a priori the occurrence of such scenarios.
Furthermore the physical relevance of these models from an astrophysical
perspective depends on many factors such as the choice of the boundary (that
for the dust model is completely arbitrary) and of the total mass of the cloud.
Stellar mass black holes may form in a matter of seconds from the complete
collapse of the core of a progenitor star with mass above 20 solar masses,
while supermassive black holes may form from clouds of up to $10^9$ solar
masses and involve much longer time scales.
These values may put some constraints on the physically allowed choices of the otherwise
free parameters $r_b$, $M_i$, $b_i$ (with $i=0,2$)
and therefore on the possibility of having the singularity globally visible
\cite{jhingan}.

At present answering the question whether these models have any
importance for realistic collapse is not possible. Different attitudes
are then possible. One could believe that Cosmic Censorship must hold, and
therefore during collapse several mechanisms must come into
play in order to form a black hole.
These `mechanisms' could be either entirely classical or of quantum-gravitational nature.
On the other hand one could believe that this kind of naked singularities
are theoretically possible and therefore ask the question of what kind of implications they
may have for astrophysics.
It could be argued that quantum corrections may resolve the singularity
in the strong field regime and that effects occurring in the ultradense region
could then propagate until the boundary thus changing completely the classical picture
\cite{QG}.
Or it could be that these effects remain confined in the close vicinity of the
center thus having no significant influence of the evolution of the outer shells
\cite{Bambi}.

In any case, despite the fact that the possible detection of effects related to the visibility of the high
density region might be very difficult since many other factors come into play, we believe that
these results enforce the idea that solutions containing naked singularities must be studied carefully
to understand whether in principle it will be possible in the future to detect some signature
of new physics coming from astrophysical events.

{\bf Aknowledgements:} The authors would like to thank Prof. A. Krasinki for useful comments and suggestions.

\appendix
\section*{Appendix}
\renewcommand{\thesection}{A}

From the integration of Eq.~\eqref{motion} we obtain $t(r,R)$, which, once inverted, gives the desired solution $R(r,t)$.
We can write $R$ in parametric form as a function of a parameter $\eta(r,t)$ in the three cases as:
\begin{equation}\label{cases}
    R(r,t) = \begin{cases}
   \frac{F}{2f}(\cosh\eta-1) \; \text{with} \; (\sinh\eta-\eta)=\frac{2(t-a)f^{\frac{3}{2}}}{F} \; ,
   \hfill \text{for $f>0$} \; , \\
  \left(\frac{3\sqrt{F}(t-a)}{2}\right)^{\frac{2}{3}} \; ,
   \hfill \text{for $f=0$} \; ,\\
  \frac{F}{2(-f)}(1-\cos\eta) \; \text{with} \;  (\eta-\sin\eta)=\frac{2(t-a)(-f)^{\frac{3}{2}}}{F}\; ,
   \hfill \text{for $f<0$} \; ,
\end{cases}
\end{equation}
with $a(r)$ being a function coming from the integration that has to be determined once the initial conditions are imposed.
On the other hand we can also integrate Eq.~\eqref{motion} to obtain $t(r,R)$. In the flat region given by $b=0$ we get
        \begin{equation}
           t(r,R)= -\frac{2R^{\frac{3}{2}}}{3\sqrt{F}}+a(r) \; .
        \end{equation}
        Once we impose the initial condition $R(r,t_i)=r$, with $t_i=0$ we get
        \begin{equation}\label{a0}
           a(r)= \frac{2r^{\frac{3}{2}}}{3\sqrt{F}}=\frac{2}{3\sqrt{M}}=t_s(r) \; .
        \end{equation}
 We define $X(r)=\pm\frac{F}{f}$ with plus sign in the hyperbolic region, given by $f>0$, and minus sign in the elliptic region, given by $f<0$ and obtain
\begin{equation}
    t(r,R) = \begin{cases}
   \frac{R}{\sqrt{f}}\left(\frac{X}{R} \tanh^{-1}\frac{1}{\sqrt{\frac{X}{R}+1}}-\sqrt{\frac{X}{R}+1}\right)+a(r) \; ,
    \text{for $f>0$} \; , \\
  \frac{R}{\sqrt{-f}}\left(\sqrt{\frac{X}{R}-1}-\frac{X}{R}\tan^{-1}\frac{1}{\sqrt{\frac{X}{R}-1}}\right)+a(r) \; ,
    \text{for $f<0$} \; .
\end{cases}
\end{equation}
   The same results in terms of the rescaled functions hold provided that we substitute $R$, $F$ and $f$ with $v$, $M$ and $b$.
The singularity curve $t_s(r)$ is given by imposing the initial condition $R(r,0)=r$ and becomes
\begin{equation}\label{a}
    a(r)=t_s(r) = \begin{cases}
   \frac{r}{\sqrt{f}}\left(\sqrt{\frac{X}{r}+1}-\frac{X}{r} \tanh^{-1}\frac{1}{\sqrt{\frac{X}{r}+1}}\right) \; ,
   \text{for $f>0$} \; , \\
  \frac{r}{\sqrt{-f}}\left(\frac{X}{r} \tan^{-1}\frac{1}{\sqrt{\frac{X}{r}-1}}-\sqrt{\frac{X}{r}-1}\right) \; ,
    \text{for $f<0$} \; .
\end{cases}
\end{equation}
The apparent horizon curve is finally given by $t_{ah}(r) = t(r,r^2M)$ and it becomes
\begin{equation}
    t_{ah}(r) = \begin{cases}
  t_s(r)+\frac{F}{f^{\frac{3}{2}}}\tanh^{-1}\sqrt{\frac{f}{1+f}}-\frac{F}{f}\sqrt{1+f} \; ,
   \hfill \text{for $f>0$} \; , \\
  t_s(r)-\frac{2}{3}F(r) \; ,
   \hfill \text{for $f=0$} \; ,\\
  t_s(r)+\frac{F}{(-f)^{\frac{3}{2}}}\tan^{-1}\sqrt{-\frac{f}{1+f}}-\frac{F}{f}\sqrt{1+f} \; ,
   \hfill \text{for $f<0$} \; .
\end{cases}
\end{equation}

\end{document}